\pdfoutput=1
\UseRawInputEncoding
\documentclass{article}
\usepackage{spconf,amsmath,graphicx}
\usepackage{multirow}
\usepackage{booktabs}
\usepackage{cite}
\usepackage[urlcolor=blue]{hyperref}
\usepackage{url}
\usepackage{bm}
\usepackage{graphicx}
\usepackage{subfigure}

\title{Enhancing the Vocal Range of Single-Speaker Singing Voice Synthesis with Melody-Unsupervised Pre-training}
\name{Shaohuan Zhou$^{1,\dagger}$\thanks{$^{\dagger}$This work was done when Shaohuan Zhou was an intern at ARC Lab, Tencent PCG.}, Xu Li$^{2,*}$, Zhiyong Wu$^{1,3,*}$\thanks{* Corresponding authors.}, Ying Shan$^2$, Helen Meng$^{1,3}$}
\address{
    $^1$ Shenzhen International Graduate School, Tsinghua University, Shenzhen, China\\
    $^2$ ARC Lab, Tencent PCG\\
    $^3$ The Chinese University of Hong Kong, Hong Kong SAR, China\\
    \small{
        zhoush21@mails.tsinghua.edu.cn, \{nelsonxli, yingsshan\}@tencent.com, zywu@sz.tsinghua.edu.cn, hmmeng@se.cuhk.edu.hk
    }
}

\ninept
\begin{document}
%
\maketitle
\begin{abstract}
The single-speaker singing voice synthesis (SVS) usually underperforms at pitch values that are out of the singer's vocal range or associated with limited training samples.
Based on our previous work, this work proposes a melody-unsupervised multi-speaker pre-training method conducted on a multi-singer dataset to enhance the vocal range of the single-speaker, while not degrading the timbre similarity.
This pre-training method can be deployed to a large-scale multi-singer dataset, which only contains audio-and-lyrics pairs without phonemic timing information and pitch annotation.
Specifically, in the pre-training step, we design a phoneme predictor to produce the frame-level phoneme probability vectors as the phonemic timing information and a speaker encoder to model the timbre variations of different singers, and directly estimate the frame-level f0 values from the audio to provide the pitch information. These pre-trained model parameters are delivered into the fine-tuning step as prior knowledge to enhance the single speaker's vocal range.
Moreover, this work also contributes to improving the sound quality and rhythm naturalness of the synthesized singing voices. It is the first to introduce a differentiable duration regulator to improve the rhythm naturalness of the synthesized voice, and a bi-directional flow model to improve the sound quality.
Experimental results verify that the proposed SVS system outperforms the baseline on both sound quality and naturalness.


\end{abstract}

\section{Introduction}
\label{sec:intro}



Singing voice synthesis (SVS) aims at synthesizing human-like singing voices given musical scores.
Unlike text-to-speech (TTS), which only considers text as input, SVS needs to generate voice with accurate pronunciation, pitch, and tempo according to the musical scores \cite{hono2018recent,gu2021bytesing}. As a subtopic of speech synthesis, SVS has been paid more and more attention due to its potential applications in virtual singers, entertainment, etc.


With recent success in deep learning techniques, singing voice synthesis has made great progress in improving sound quality \cite{lu2020xiaoicesing,nishimura2016singing,chen2020hifisinger}.
Ren et al. \cite{ren2020deepsinger} developed a pipeline to synthesize singing voices with data mined from the web. Chen et al. \cite{huang2022singgan} proposed SingGAN which is specifically designed for singing voice vocoding. Other attempts \cite{zhang2021visinger,zhou22f_interspeech} were proposed to train the SVS system in an end-to-end manner to alleviate accumulated errors.
Human singing voices are usually limited by the singer's vocal range because they cannot produce pitch values out of their vocal ranges.
Similarly, the single-singer SVS system has the same problem because the training data is limited to a specific vocal range as well.
Moreover, the single-singer SVS system also fails at pitch values that are associated with limited training samples, even if they are within the singer's vocal range.
To extend the vocal range of the SVS model, \cite{zhang2022wesinger} tried pitch shifting as a data augmentation method. 
The pitch of each song is either increased or decreased by one semitone to enlarge the vocal range of the training data.
However, they claimed that this method could lead to somewhat perceptible changes in timbre.

To explore the potential ability to develop virtual singers with wide vocal ranges, we extend our previous work \cite{zhou22f_interspeech} with a melody-unsupervised pre-training method conducted on a large-scale multi-singer dataset to enhance the vocal range of a single-speaker SVS system, while not degrading the timbre similarity.
This pre-training method can be deployed to datasets that only contain audio-and-lyrics pairs without phonemic timing information and pitch annotation.
Inspired by \cite{choi2022melody}, we design a phoneme predictor to produce frame-level phoneme probabilities as phonemic timing information and directly estimate the frame-level f0 values from the audio to provide the pitch information.
Besides, we make some further modifications in order to deploy it to our previous framework: 1) the input features for the phoneme predictor are latent features predicted by the posterior encoder, rather than the Mel-spectrograms in \cite{choi2022melody}; 2) we introduce an additional speaker encoder in the prior encoder to model the timbre variations of different singers, while \cite{choi2022melody} conducted experiments on the single-singer dataset.
It is also worth noting that the role of the melody-unsupervised training in this work is as a pre-training strategy to enhance the single speaker's vocal range, while \cite{choi2022melody} merely aims at exploring the potential feasibility of melody-unsupervised SVS training.

Moreover, this work also contributes to improving the sound quality and rhythm naturalness of the synthesized voice. It is observed in our previous work \cite{zhou22f_interspeech} that some phonemes are pronounced slightly earlier or later than they should be, resulting in an unnatural-sounding rhythm. Motivated by \cite{tan2022naturalspeech}, we leverage a differentiable duration regulator and a bi-directional flow model to enhance the sound quality and rhythm naturalness. To the best of our knowledge, this work is the first to adopt the differentiable duration regulator and the bi-directional flow in the SVS systems. 

\begin{figure*}[ht]
  \centering
  \subfigure[\label{subfig:a}The multi-singer pre-training step]{\includegraphics[width=0.50\textwidth]{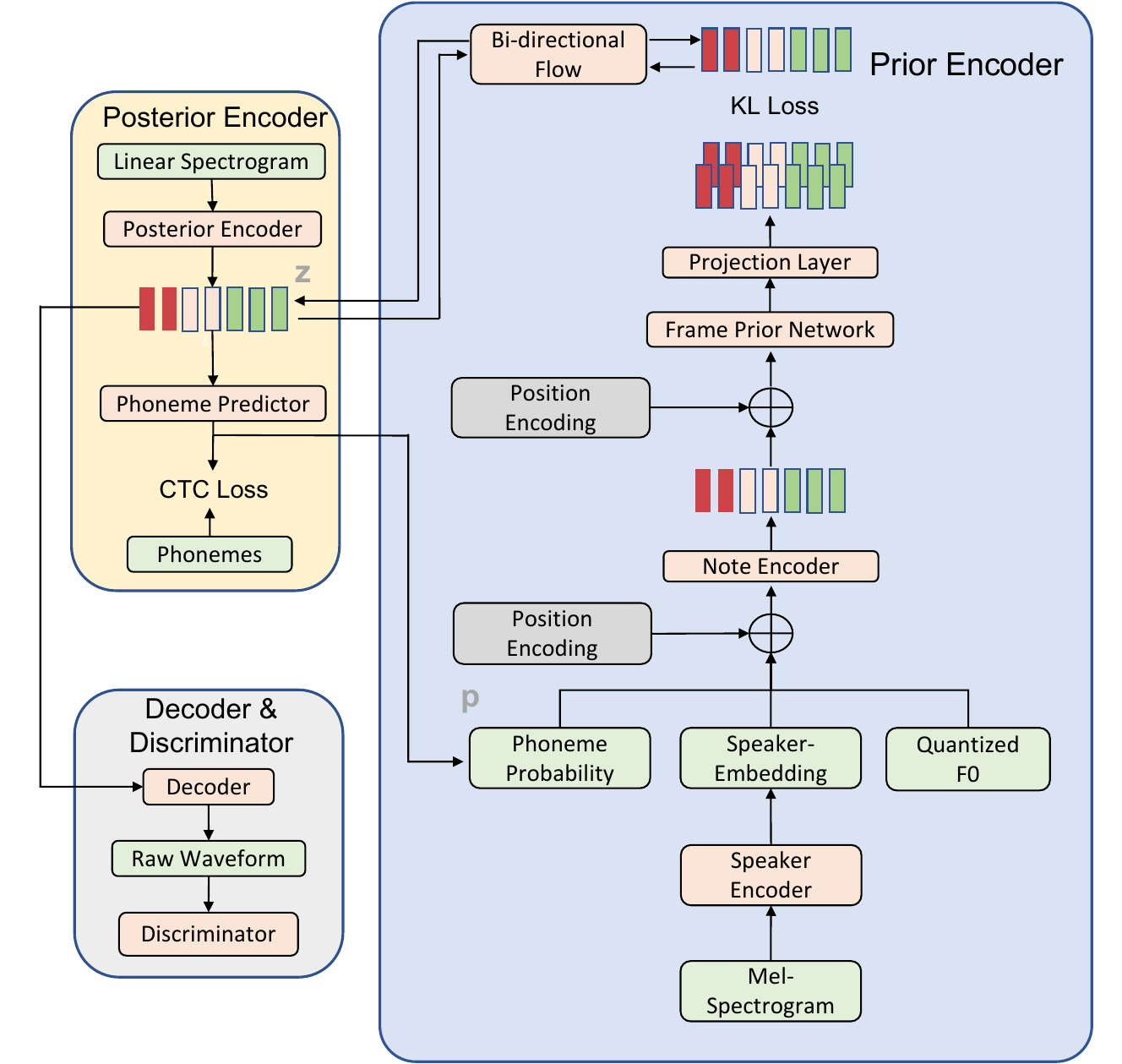}}
  \quad 
  \subfigure[\label{subfig:b}The single-singer fine-tuning step]{\includegraphics[width=0.40\textwidth]{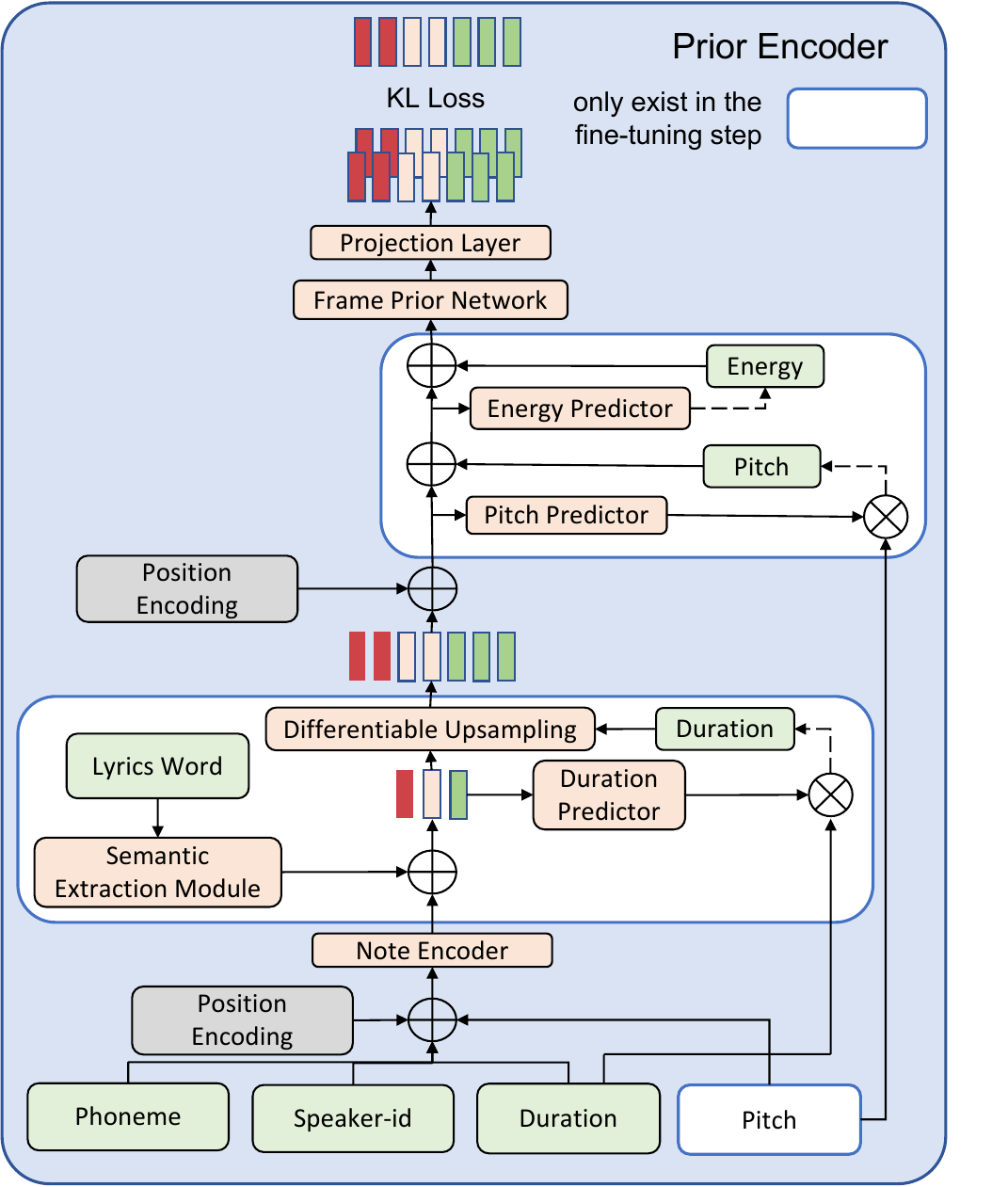}}
  \caption{The structure of the proposed SVS system, where (a) represents the pre-training stage, while (b) represents the fine-tuning stage. Modules with blue boxes and highlighted backgrounds only exist in the fine-tuning step.}
  \label{fig: architecture}
\end{figure*}


The contributions of this work are summarized as follows: 1) Proposing a melody-unsupervised multi-speaker pre-training strategy to enhance the vocal range of the single-singer SVS system; 2) Exploring the potential feasibility of developing virtual singers with wide vocal ranges; 3) Introducing a differentiable up-sampling layer and a bi-directional flow model to improve the sound quality and rhythm naturalness. Some audio samples are provided for listening\footnote{Sample: \href{https://thuhcsi.github.io/melody-unsupervised-pretraining-svs/}{https://thuhcsi.github.io/melody-unsupervised-pretraining-svs}}.

The rest of this paper is organized as follows: Section~\ref{sec:Method} illustrates the proposed system. Experimental setup and experiment results are demonstrated in Section~\ref{sec:expt-setup} and \ref{sec:expt-rst}, respectively. We conclude this work in Section~\ref{sec:conclusion}.



\section{The Proposed Method}
\label{sec:Method}
The training stage of the proposed model consists of two steps: the multi-singer pre-training step and the single-singer fine-tuning step. The architecture of the proposed model is illustrated in Fig.\ref{fig: architecture}, which consists of a prior encoder, a posterior encoder, and a decoder together with a discriminator.
The proposed model is designed from our previous work \cite{zhou22f_interspeech} with the following modifications.
The posterior encoder utilizes a phoneme predictor to predict frame-level phoneme probabilities in the pre-training step. 
The prior encoder adds a speaker encoder to model the timbre variations, replaces the length regulator with a differentiable duration regulator to improve the rhythm naturalness, and upgrades the flow module to be bi-directional to improve the sound quality.




\subsection{The Melody-Unsupervised Multi-Singer Pre-Training Step}
Since the multi-singer training data has no phonemic timing information, in the pre-training step, this work utilizes the automatic speech recognition (ASR) training strategy to train a phoneme predictor in the posterior encoder and predict the frame-level phoneme probabilities $p$.
The probability vectors are multiplied with the phoneme look-up table to obtain the frame-level phoneme embeddings.
In addition, the continuous pitch $f_{0}$ is estimated from the audio and quantized into the note pitch.
The note pitch is passed through the embedding layer to obtain frame-level pitch embeddings.
Moreover, we apply a speaker encoder to extract frame-level speaker embeddings to model the timbre variations of different singers.
Since the pre-training step directly deals with estimated pitch values and focuses on enhancing the vocal range, the pitch predictor, the energy predictor and the duration-related modules are dropped during the pre-training step.

\subsubsection{Phoneme predictor}
We train the phoneme predictor using the connectionist temporal classification (CTC) \cite{graves2006connectionist} loss. 
It contains two layers of FFT blocks and one linear layer.
The linear layer maps the hidden channels to the number of phoneme categories. 
We obtain the probability vector $p$ after taking the softmax operation on the linear layer's output, then multiply it with the phoneme look-up table to get frame-level phoneme embeddings.
Meanwhile, we take the log function of $p$ to compute the CTC loss with the ground truth phoneme sequences. 

\subsubsection{Speaker encoder}
This work adopts one of the state-of-the-art speaker recognition models, i.e. ECAPA-TDNN \cite{desplanques2020ecapa}, as the speaker encoder. Its advanced network architecture and attentive statistics pooling layer have shown great effectiveness in both speaker recognition \cite{desplanques2020ecapa} and voice conversion \cite{guo2022improving,li2022hierarchical}. The speaker encoder is configured as the one with 512 channels in Table 1 of \cite{desplanques2020ecapa}, and it extracts 192-dimensional frame-level speaker embeddings from the audio's Mel-Spectrograms. These embeddings are given as the speaker condition in the multi-singer pre-training step. 

\subsection{The Single-Singer Fine-Tuning Step}
In the fine-tuning step, it loads the pre-trained model parameters, then utilizes the single-singer Opencpop dataset to fine-tune model parameters.
It uses the phoneme and note-pitch annotations provided by the dataset to derive the phoneme and pitch embeddings, instead of using the phoneme probability vectors and quantized f0 values in the pre-training step.
Note that the phoneme and note-pitch annotations are at the phoneme level, rather than the frame level, such that a duration regulator after the note encoder is necessary to up-sample the embedding vectors into the frame level.
As for the speaker embedding, we use the pre-trained speaker encoder to extract an averaged speaker embedding over the Opencpop dataset, then utilize it as a fixed speaker condition during the fine-tuning step.
Moreover, the energy predictor and the pitch predictor join the fine-tuning process to enhance the expressiveness and pitch accurateness of the synthesized samples, following our previous work\cite{zhou22f_interspeech}.
The synthesizable vocal range of the model is enhanced due to the multi-singer pre-training on a large-scale dataset.

\subsection{Differentiable Duration Regulator}
Most previous SVS systems simply replicate each phoneme hidden representation with the predicted duration in a hard way, which may degrade the rhythm naturalness.
Inspired by \cite{tan2022naturalspeech}, we leverage a differentiable duration regulator, which contains a duration predictor and a differentiable up-sampling layer. 
The duration predictor outputs the ratio of each phoneme to the corresponding note duration, then the ratio is multiplied by the note duration and fed to the differentiable up-sampling layer.
The differentiable up-sampling layer leverages the predicted duration to learn a projection matrix to extend the phoneme hidden sequence from the phoneme level to the frame level.
It makes the phoneme-to-frame expansion differentiable and thus can be jointly optimized with other modules in the system.

\subsection{Bi-directional Flow}
In the previous work \cite{zhou22f_interspeech}, the flow model maps the complex posterior distribution to the simple prior distribution in the training stage while operating reversely in the inference stage.
This process suffers from the mismatch problem between the training and inference stages.
Therefore, we leverage a bi-directional flow module\cite{tan2022naturalspeech} during training, which bridges the complex posterior distribution and the simple prior distribution bi-directionally to alleviate the mismatch issue in the inference stage.
It is worth noting that we observed that the system can easily fail to train and encounter gradient explosion when the KL losses on both sides of the flow contribute equally, so we define the reverse KL loss weight as 0.5.

\section{Experimental Setup}
\label{sec:expt-setup}

\subsection{Datasets}
%
The experiments are conducted on two open-sourced Mandarin singing corpus, named OpenSinger \cite{huang2021multi} and Opencpop \cite{wang2022opencpop}. The OpenSinger dataset is used for pre-training, it contains 50 hours of pop songs, including 30 hours from 41 females and 20 hours from 25 males. The singing data is saved in the WAV format, sampled at 24 kHz, and quantized by 16 bits.
The Opencpop dataset is used for fine-tuning and testing the system. 
It consists of 100 popular Mandarin songs sampled at an audio rate of 44.1 kHz and recorded by a female professional singer. The recordings have been annotated with phoneme, note pitch, note duration and phonetic timing information. All audio is down-sampled to 24 kHz with 16-bit quantization for experiments. The whole dataset consists of 3,756 audio clips, with a total duration of around 5.2 hours. 3550 clips of them are selected for training while the other 206 utterances are for testing.

\begin{table*}[th]
\normalsize
\renewcommand{\arraystretch}{1.2}
  \caption{Experimental results in terms of subjective mean opinion score (MOS) with 95\% confidence intervals and four objective metrics.}
  \setlength{\tabcolsep}{2mm}{
  \label{tab:mos}
  \centering
  \begin{tabular}{l|cccccc} 
    \toprule
    \textbf{Model} &\textbf{Sound quality}&\textbf{Naturalness}&\textbf{F0 MAE $\downarrow$} & \textbf{F0 Correlation (Hz) $\uparrow$}& \textbf{Duratin MAE $\downarrow$} & \textbf{Speaker Similarity $\uparrow$}\\
    \midrule
    Baseline & $3.350\pm0.103$& $3.317\pm0.104$& $6.207~(11.270)$ &$0.984~(0.972)$&$2.262$&$0.862$  \\
    Proposed & \bm{$3.708\pm0.099$}& \bm{$3.667\pm0.102$}& \bm{$6.051~(8.118)$}&\bm{$0.986~(0.984)$}&\bm{$2.086$}&\bm{$0.884$}  \\
    \midrule
    Recording & $4.854\pm0.058$ & $4.833\pm0.057$ & $-$ & $-$ & $-$ & $-$\\
    \bottomrule
  \end{tabular}}
\end{table*}

\begin{figure*}[htbp]
	\centering
	\subfigure[Ground truth] {\includegraphics[width=.28\textwidth]{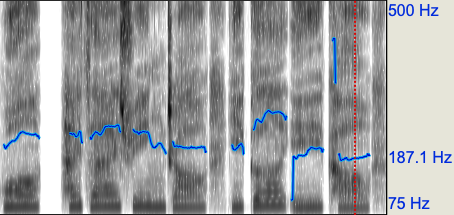}}
	\subfigure[Proposed] {\includegraphics[width=.28\textwidth]{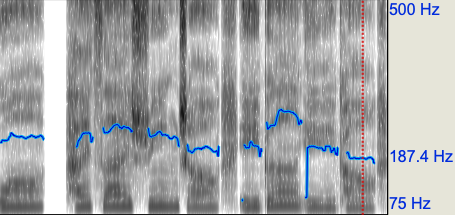}}
	\subfigure[Baseline] {\includegraphics[width=.28\textwidth]{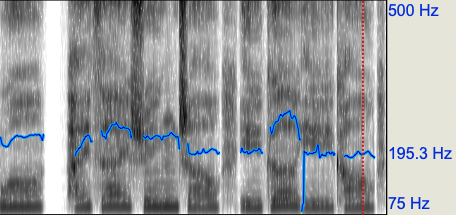}}
	\caption{Visualizations of spectrograms and pitch contours in three systems: Ground truth, Proposed, and Baseline. The blue line represents the pitch contour. The numbers on the right represent the range of pitches (Hz) and the pitch value at the red line.}
	\label{case}
\end{figure*}
\subsection{System Configuration}
Two systems are constructed for evaluating the effectiveness of the proposed method.

\textbf{Baseline:}
The baseline model follows the structure of the VISinger system \cite{zhang2021visinger} and also adopts the improvements proposed by our previous work \cite{zhou22f_interspeech}.
The dimension of pitch, phoneme, and duration embedding is 192.
Each note pitch is converted into a pitch ID following the MIDI standard\cite{midi}, while the note duration is converted into the number of frames. 
The dimension of text embedding derived from BERT is 768 and converted into 192 by an additional linear layer in the text encoder. 
The note encoder contains 6 FFT blocks, and the duration predictor consists of 3 one-dimensional convolutional networks.
Moreover, the pitch and energy predictors are designed to predict the pitch and energy of each frame, respectively. They are configured as \cite{zhou22f_interspeech} and output 192-dimensional embedding vectors.

\textbf{Proposed:}
The proposed method adopts all the contributions aforementioned.
The phoneme vocabulary size is 61.
We pre-processed the phonemes in OpenSinger since they differ from those in the Opencpop dataset.
Specifically, we re-extract the phonemes from the transcripts provided by the dataset using the ``pypinyin'' algorithm.\footnote{Source-code: \href{https://github.com/mozillazg/python-pinyin}{https://github.com/mozillazg/python-pinyin}}
The other hyperparameter settings are consistent with those in the baseline.

The proposed method is firstly pre-trained on OpenSinger with 100k steps, then fine-tuned by 200k steps on Opencpop. The baseline model is trained up to 300k steps on Opencpop. 
We adopt the AdamW\cite{loshchilov2017decoupled} optimizer with $\beta_1$ = 0.8, $\beta_2$ = 0.99, $\epsilon$ = $10^{-9}$. The initial learning rate is set to $1e^{-4}$, with a learning decay of $0.999875$.
All the models are trained on 2 Nvidia Tesla A100 devices, and the batch size on each GPU is 8.

\section{Experiment Results}
\label{sec:expt-rst}

\subsection{Performance Comparison}
The mean opinion score (MOS) test is conducted to evaluate the quality and naturalness of the synthesized singing voices.
For each system, we randomly select 10 audio samples, with each sample within about 10 seconds. 
24 Chinese native speakers are recruited to rate the testing samples by a score from 1 to 5 (the higher, the better), with a 1-point interval. 
As shown in Table.\ref{tab:mos}, our method achieves a better MOS of 3.708 in sound quality and 3.667 in naturalness, exceeding baseline by 0.358 and 0.350.
It demonstrates that the proposed method can improve the quality and naturalness of the synthesized singing voice effectively.

We further calculate some objective metrics including F0 mean absolute error (F0 MAE), duration mean absolute error (Duration MAE), and F0 correlation\cite{lu2020xiaoicesing}.
These three metrics represent the accuracy of pitch and duration prediction.
As shown in Table.\ref{tab:mos}, our proposed method outperforms the baseline in all three metrics mentioned above. 
This demonstrates that the use of the melody-unsupervision pre-training method to expand the vocal range and the use of differentiable up-sampling layer can lead to more accurate predictions of pitch and duration, respectively

Moreover, we also calculate the speaker similarity between synthesized and ground-truth waveforms.
We first apply the pre-trained speaker recognition model released in\cite{jia2018transfer} to extract the speaker embedding of the waveform.
Then the cosine similarity is performed between the speaker embeddings of the synthesized singing voice and the ground-truth waveform, which reflects the speaker similarity. 
As can be seen in Table.\ref{tab:mos}, the proposed method achieves better speaker similarity compared with the baseline, showing that using speaker encoder to model timbre information can improve the timbre similarity of synthesized singing voice.

To further verify the effectiveness in extending the pitch range, we selected audio segments from the testing set that contained pitches below F3 or above D5 and recalculated the F0 metrics.
As shown in the bracketed data of `F0 MAE' and `F0 Correlation' in Table.\ref{tab:mos}, the proposed method has a much larger gap over the baseline method when synthesizing high- or low-pitched singing voice, verifying the validity of our contribution.

\subsection{Ablation Study}
To demonstrate the effectiveness of different contributions in the proposed method, we carry out three ablation studies.
The comparison mean opinion score (CMOS) is used to compare the synthesized singing voices in terms of both naturalness and quality.
The naturalness and quality are evaluated by a single CMOS score to reflect the overall impact of each proposed contribution and also save labor efforts.
As shown in Table.\ref{tab:cmos}, the absence of pre-training, bi-direction flow and differentiable duration regulator results in -0.857 CMOS, -0.335 CMOS and -0.143 CMOS, respectively.
This demonstrates the validity of the above three contributions in improving the quality and naturalness of synthesized singing voice.
The audio used for comparison in the ablation study can be heard on our demo page.

\begin{table}[th]
\renewcommand{\arraystretch}{1.1}
  \caption{Ablation study results.}
  \begin{center}

  \setlength{\tabcolsep}{4mm}{
  \label{tab:cmos}
  \centering
  \begin{tabular}{l|c} 
    \toprule
    \textbf{Model} &\textbf{CMOS} \\
    \midrule
    Proposed & $0.000$ ~~~\\
     - pretrain & - $0.857$ ~~~  \\
     - bi-direction flow & - $0.335$ ~~~  \\
     - differentiable duration regulator& - $0.143$ ~~~  \\
    \bottomrule
  \end{tabular}}
        
  \end{center}
\end{table}

\subsection{Case Study}
To demonstrate the impact of the aforementioned contributions, a case study is conducted to synthesize a testing sample that contains pitch values of limited training data.
We compare the ground-truth, the proposed method and the baseline.
As shown in \ref{case}, the pitch is marked with blue lines and the pitch value at the red line is shown on the right. 
This sample ends with a slightly low pitch that is associated with few training data. 
It is observed that the proposed method synthesizes this pitch accurately, but the baseline method tends to incorrectly use a higher pitch to replace this one, proving that the proposed pre-training strategy is effective in enhancing the vocal range.

\section{Conclusion}

This paper proposes a melody-unsupervised pre-training method conducted on a multi-singer dataset to enhance the vocal range of the single-speaker, while not degrading the timbre similarity.
Moreover, it also contributes to improving the sound quality and rhythm naturalness of the synthesized singing voices. It is the first to introduce a differentiable duration regulator to improve the rhythm naturalness, and a bi-directional flow model to improve the sound quality.
Experiments on Opencpop show that the proposed method outperforms the baseline in both subjective and objective evaluations.


\section{Acknowledgement}
\label{sec:conclusion}
This work is supported by National Natural Science Foundation of China (62076144), Shenzhen Key Laboratory of next generation interactive media innovative technology (ZDSYS20210623092001004),\\ Shenzhen Science and Technology Program (WDZC2022081614051\\5001) and Tencent AI Lab Rhino-Bird Focused Research Program (RBFR2022005).




\bibliographystyle{IEEEbib}
\bibliography{strings,refs}

\end{document}